\documentclass[11pt]{article}
\usepackage[utf8]{inputenc}
\usepackage{cancel}
\usepackage{amssymb,amsmath,mathrsfs,enumerate}
\usepackage{graphicx,rotate,multicol}
\usepackage{float}
\usepackage{tocloft}
\usepackage{subfig}
\usepackage[margin=10pt,labelfont=bf]{caption}
\usepackage{cite}
\usepackage{soul}
\usepackage{mathabx}
\usepackage{color}
\usepackage[colorlinks=true,
            linkcolor=blue,
            urlcolor=blue,
            citecolor=red]{hyperref}
\usepackage{eso-pic}
\usepackage{lipsum}

\makeatletter
\let\Hy@linktoc\Hy@linktoc@page
\makeatother

\definecolor{ourcolor}{rgb}{0.7, 0.25, 0.05}

\let\tilde=\widetilde

\let\bar=\overline

\def \order(#1){{\mathcal O} \left(#1 \right)}

\evensidemargin=\oddsidemargin
\newcommand{\ba}{\begin{array}}
\newcommand{\ea}{\end{array}}
\newcommand{\bd}{\begin{displaymath}}
\newcommand{\ed}{\end{displaymath}}
\newcommand{\besub}{\begin{subequations}}
\newcommand{\eesub}{\end{subequations}}
\newcommand{\bea}{\begin{eqnarray}}
\newcommand{\eea}{\end{eqnarray}}

\allowdisplaybreaks

\title{\color{black}{\bf Connecting ANITA Anomalous Events to a Non-thermal Dark Matter Scenario
}}

\author {\bf   Debasish Borah,$^{a}$\footnote{dborah@iitg.ac.in}
\hspace{4pt}   Arnab Dasgupta,$^{b}$\footnote{arnabdasgupta@protonmail.ch}
\hspace{4pt}   Ujjal Kumar Dey,$^{c,d}$\footnote{ujjal@apctp.org, ujjal@iiserbpr.ac.in} 
\hspace{4pt}   Gaurav Tomar$^{e}$\footnote{tomar@sogang.ac.kr} \\[10pt]
\small\em $^a$Department of Physics, 
		      Indian Institute of Technology Guwahati,
		      Assam 781039, India \\
\small\em $^b$School of Liberal Arts, 
              Seoul-Tech, 
              Seoul 139-743, Korea \\
\small\em $^c$Asia Pacific Center for Theoretical Physics,
              Pohang 37673, Korea\\
\small\em $^d$Department of Physical Sciences, 
              Indian Institute of Science Education 
              and Research Berhampur,\\ 
              \small\em Transit Campus, Government ITI, 
              Berhampur 760010, Odisha, India\\              
\small\em $^e$Department of Physics, 
              Sogang University, 
              Seoul, Korea, 121-742
}

\date{}

\begin{document}



\maketitle

\begin{abstract}
The ANtarctic Impulsive Transient Antenna (ANITA) collaboration has observed two EeV-energy, upward going events originating from below the horizon. As no standard model (SM) particles, propagating through the earth at such energy and exit angles, can give rise to the required survival probability for the observed events, several beyond standard model (BSM) proposals have come up. We propose a scenario where a $Z_2$ odd sector is responsible for such anomalous events. The next to lightest $Z_2$ odd particle or the next to lightest stable particle (NLSP), created from ultra high energy neutrino interactions with nucleons, can pass through the earth and then decay into the lightest $Z_2$ odd particle or the lightest stable particle (LSP) and a tau lepton. The long lived nature of the NLSP requires its coupling with the LSP to be very small, keeping the LSP out of thermal equilibrium in the early universe. The LSP can then be a non-thermal dark matter while the tau leptons produced from NLSP decay after passing through earth can explain the ANITA events. We first show that a purely non-thermal dark matter scenario can not give rise to the required event rates while a hybrid scenario where dark matter can have both thermal as well non-thermal contribution to its relic abundance, serves the purpose.
\end{abstract}


\newpage


\section{Introduction}

Recently, the ANtarctic Impulsive Transient Antenna (ANITA) collaboration has reported two anomalous upward going ultra high energy cosmic ray (UHECR) air shower events corresponding to deposited energies of $0.6\pm 0.4$ EeV and $0.56^{+0.3}_{-0.2}$ EeV respectively \cite{Gorham:2016zah, Gorham:2018ydl}. These two events, which originate from well below the horizon, with angles of elevation $-27.4^{\circ} \pm 0.3^{\circ}$ and  $-35.0^{\circ} \pm 0.3^{\circ}$ respectively, have $\geq 3\sigma$ evidence for occurring due to direct upward moving earth emergent UHECR like air showers above the Antarctic ice surface \cite{Gorham:2018ydl}. However, no particle in the standard model (SM) can survive such a passage through earth, a distance of several thousand kilometers corresponding to the observed zenith angles for ANITA anomalous events (AAEs). Therefore, several beyond standard model (BSM) proposals have been put forward to explain the observed events. They include sterile neutrino mixing \cite{Cherry:2018rxj, Huang:2018als}, heavy dark matter (DM) \cite{Anchordoqui:2018ucj, Dudas:2018npp, Heurtier:2019git, Hooper:2019ytr, Chipman:2019vjm, Heurtier:2019eou, Yin:2018yjn, Yin:2017wxm}, inelastic boosted DM~\cite{Cline:2019snp, Heurtier2019}, long lived charged particles like stau \cite{Fox:2018syq, Connolly:2018ewv}, R-parity violating supersymmetry (SUSY)~\cite{Collins:2018jpg}, SUSY sphalerons configurations \cite{Anchordoqui:2018ssd}, leptoquarks~\cite{Chauhan:2018lnq}, and radio pulses from axion-photon conversion~\cite{Esteban2019}. Additionally, in Ref.~\cite{Romero-Wolf2019}, $\nu_\tau$ neutrino origin of AAEs is studied in compliance with Auger~\cite{Aab2015} and IceCube~\cite{Aartsen2016} upper limits. Note that explanations of AAEs in SM scenario are also proposed~\cite{Vries2019, Shoemaker2019}, but they remain speculative and require experimental confirmations.

Here we propose a specific type of dark matter scenario where DM relic abundance is generated partially or fully from a non-thermal mechanism. DM contributes around $27\%$ to the present universe's energy density according to the latest cosmology data provided by the Planck satellite \cite{Aghanim:2018eyx}. In terms of density 
parameter $\Omega_{\rm DM}$ and $h = \text{Hubble Parameter}/(100 \;\text{km} ~\text{s}^{-1} 
\text{Mpc}^{-1})$, the present DM abundance is conventionally reported as \cite{Aghanim:2018eyx}:
$\Omega_{\text{DM}} h^2 = 0.120\pm 0.001$
at 68\% CL. Since none of the SM particles could provide a suitable DM candidate, several BSM proposals have come up out of which the weakly interacting massive particle (WIMP) paradigm has been the most popular one. In this framework, a dark matter candidate typically with electroweak scale mass and interaction rate similar to electroweak interactions can give rise to the correct dark matter relic abundance, a remarkable coincidence often referred to as the \textit{WIMP Miracle}. However, none of the experiments looking for direct detection of DM has found any signal yet, motivating the community to seek a paradigm shift. One such scenario that has drawn attention in the last few years is non-thermal dark matter \cite{Hall:2009bx}. In such a framework, DM particles have so feeble interactions with the remaining thermal bath that it never attains thermal equilibrium at any epoch in the early universe. However, it can be produced from decay of some heavy particles or scattering processes, popularly known as the freeze-in mechanism \cite{Hall:2009bx, Konig:2016dzg, Biswas:2016bfo, Biswas:2016iyh}, leading to a new paradigm called freeze-in (or feebly interacting) massive particle (FIMP). For a recent review of this DM paradigm, please see \cite{Bernal:2017kxu}. Typical FIMP or non-thermal DM models involve tiny couplings of DM with SM or other particles in the thermal bath and often lead to long lived particles. Such long lived charged particles have the potential to pass through the earth before decaying into DM and a tau lepton which could explain the anomalous events observed by ANITA~\cite{Fox:2018syq, Connolly:2018ewv}.

Motivated by these, we propose a model which is an extension of the scotogenic model \cite{Ma:2006km} that provides a common origin of DM and light neutrino masses. The original scotogenic model is an extension of the SM by three right handed neutrinos and one additional scalar doublet all of which are odd under an in-built and unbroken $Z_2$ symmetry. It is worth mentioning at this point that the observation of non-zero neutrino masses and large leptonic mixing \cite{Tanabashi:2018oca} has also been another motivation for BSM physics for last few decades. While the addition of singlet right handed neutrinos to the SM content can give rise to the usual seesaw mechanism \cite{Minkowski:1977sc, Mohapatra:1979ia, Schechter:1980gr} for neutrino mass at tree level, the scotogenic framework can explain the origin of neutrino mass and dark matter in a unified manner. The $Z_2$ odd particles, the lightest of which is the DM candidate, give rise to light neutrino masses at one loop level in this model. We extend this model suitably to explain the AAEs. We first consider an extension of this model to incorporate a purely non-thermal or FIMP type DM and name it as Model-I. Here, the DM is a gauge singlet right handed neutrino whose relic is generated purely from the non-thermal contribution, by virtue of it's small couplings to SM leptons. Although correct DM properties as well as light neutrino masses can be realized in such a scenario, it fails to explain the AAEs. We then consider a hybrid setup where DM relic receives both thermal as well as non-thermal contributions. In this scenario, referred to as Model-II hereafter, DM is the lightest neutral component of the $Z_2$ odd scalar doublet. Due to electroweak gauge interactions, the DM in this model can not be of purely non-thermal origin. However, it can receive non-thermal contribution from the decay of heavier particles after it freezes out, similar to the recent works \cite{Borah:2017dfn, Biswas:2018ybc}. In such scenarios, the thermally under-abundant DM can satisfy the correct relic abundance criteria due to the late non-thermal contributions. We show how the correct DM and neutrino phenomenology can be obtained in this hybrid model along with the explanation for the AAEs.

This paper is organised as follows. In section \ref{sbsc:model1} and \ref{sbsc:model2} we discuss the two models along with the corresponding DM phenomenology. In section \ref{sec:anitaEvts} we discuss the origin of ANITA anomalous events in both the models. We briefly comment upon different ways to probe our models in section \ref{sec:comp} and finally conclude in section \ref{sec:concl}.

\section{Model}
\label{sec:models}

\subsection{Model-I: Pure Freeze-in}
\label{sbsc:model1}
As mentioned earlier, we first discuss a model where DM relic is generated purely from the freeze-in mechanism. The particle content of Model-I are shown in table \ref{tab:a2}. There are five different types of BSM fields all of which are odd under an in-built and unbroken $Z_2$ symmetry. We need three copies of singlet right handed neutrino $N$ in order to account for non-thermal DM and light neutrino masses simultaneously. The lightest of the right handed neutrino $N_1$ is the DM candidate having tiny Yukawa couplings with leptons while the heavier right handed neutrinos generate the light neutrino masses at radiative level. While $\Phi_2, N$ are same as those in the minimal scotogenic model and are sufficient to realize $N_1$ as freeze-in DM and generate light neutrino masses and mixing, the other three types of particles $E_{L,R}, \psi_1, \psi_2$ are required in order to explain the AAEs. It is worth mentioning that due to very small Yukawa couplings associated with one of the right handed neutrinos $N_1$, the lightest neutrino mass becomes vanishingly small, predicting a hierarchical light neutrino mass spectrum.

\begin{table}[h]
\centering
\begin{tabular}{|c|cccccc|ccccc|}
\hline
 Fields & $Q$ & $u_R$ & $d_R$ & $L$  & $e_{R}$  &  $\Phi_1$ & $\Phi_2$ & $E_{L,R}$ & $N$& $\psi_1$& $\psi_2$  \\
\hline
$SU(3)$ & 3 & 3 & 3 & 1 & 1 & 1 & 1 &1 & 1 & 3 & 3 \\
\hline
$SU(2)$ & 2 & 1 & 1 & 2 & 1 & 2 & 2 &1 & 1 & 1 & 1 \\
\hline
$U(1)_Y$ & $\frac{1}{6}$ & $\frac{2}{3}$ & $-\frac{1}{3}$ & -$\frac{1}{2}$ & -1 & $\frac{1}{2}$ & $\frac{1}{2}$ & $\pm 1$ & 0 & $\frac{1}{3}$ & $\frac{4}{3}$ \\
\hline
$Z_2$ & 1 & 1 & 1& 1 & 1 & 1 & -1 &-1 & -1 & -1 & -1 \\
\hline
\end{tabular}
\caption{\label{tab:a2} Field content and transformation properties under the symmetry of the model. }
\end{table}

The relevant part of the Yukawa Lagrangian is given by,
\begin{align}
\mathcal{L}_Y & \supset Y_1 \overline{L} \Phi_1 e_R + Y_2 \overline{L} \Phi_2 E_R+Y'_2 L^T \tilde{\Phi_2} E_L + Y_3 \overline{L} \tilde{\Phi_2} N + Y_4 \overline{u_R} \psi^{\dagger}_1 E_L + Y_5 \overline{d_R} \psi^{\dagger}_2 E_L  \nonumber \\
& +Y'_4 u_R \psi_1 E_R + Y'_5 d_R \psi_2 E_R+ Y_6 d_R \psi_1 N.
\label{yukawamodel1}
\end{align} 
The scalar potential is,
\begin{align}
V(\Phi_1,\Phi_2, \psi_1, \psi_2) &=  \mu_1^2|\Phi_1|^2 +\mu_2^2|\Phi_2|^2+\frac{\lambda_1}{2}|\Phi_1|^4+\frac{\lambda_2}{2}|\Phi_2|^4+\lambda_3|\Phi_1|^2|\Phi_2|^2 \nonumber \\
& +\lambda_4|\Phi_1^\dag \Phi_2|^2 + \left\lbrace\frac{\lambda_5}{2}(\Phi_1^\dag \Phi_2)^2 + \text{h.c.}\right\rbrace + \mu^2_3 |\psi_1|^2 + \mu^2_4 |\psi_2 |^2 \nonumber \\
& + |\Phi_1|^2 ( \lambda_6 |\psi_1|^2 + \lambda_7 |\psi_2|^2) + |\Phi_2|^2 ( \lambda_8 |\psi_1|^2 + \lambda_9 |\psi_2|^2) \nonumber \\
& + \left\lbrace\frac{\lambda_{10}}{2}(\Phi_1^\dag)^2 (\psi^{\dagger}_1 \psi_2) + \text{h.c.}\right\rbrace + \left\lbrace\frac{\lambda_{11}}{2}(\Phi_2^\dag)^2 (\psi^{\dagger}_1 \psi_2) + \text{h.c.}\right\rbrace \nonumber \\
& + \lambda_{12} |\psi_1 |^4 + \lambda_{13} |\psi_2 |^4.
\label {c2}
\end{align}
As mentioned, the lightest right handed neutrino $N_1$ is the dark matter candidate in this model and being gauge singlet, it can be prevented from being produced thermally in the early universe, if the corresponding Yukawa couplings are very small. We consider such a possibility where $N_1$ is produced non-thermally from the decay of $\Phi_2$. The decay of $\Phi_2$ can occur either in equilibrium or after its freeze-out, the latter is similar to the superWIMP formalism \cite{Feng:2003uy}. It should be noted that all the components of $\Phi_2$ can contribute to the non-thermal production of $N_1$ namely, $\Phi^{\pm}_2 \rightarrow N_1 l^{\pm}, \Phi^0_2 \rightarrow N_1 \nu, \; l \equiv e, \mu, \tau$. Since the $Z_2$ symmetry remains unbroken, the neutral components of $\Phi_2$ do not acquire any non-zero vacuum expectation value (VEV) which can be ensured by choosing $\mu_2^2 >0$. We parametrize the two scalar doublets $\Phi_{1,2}$ as 
\begin{equation}
\Phi_1 =
  \begin{pmatrix}
    0\\
 \frac{1}{\sqrt 2}(v+h)
  \end{pmatrix},
\Phi_2 =
  \begin{pmatrix}
    \Phi_2^+\\
 \frac{1}{\sqrt 2}(\Phi^R_2 + i \Phi^I_2)
  \end{pmatrix},
  \end{equation}
where $\Phi_1$ is identified with the SM Higgs doublet whose neutral component acquires a VEV denoted by v, responsible for electroweak symmetry breaking (EWSB). As follows from the scalar potential, after EWSB, the physical scalars (originating from the two scalar doublets) have the following masses,
\besub
\bea
m_{h}^2 &=& \frac{1}{2}\lambda_1 v^2,  \\
(M^R_{\Phi_{2}})^2 &=& 
\mu_2^2 + \frac{1}{2}(\lambda_{3} + \lambda_{4} + \lambda_{5})v^2, \\
(M^I_{\Phi_{2}})^2 &=& 
\mu_2^2 + \frac{1}{2}(\lambda_{3} + \lambda_{4} - \lambda_{5})v^2, \\
(M^{\pm}_{\Phi_{2}})^2 &=& 
\mu_2^2 + \frac{1}{2} \lambda_{3} v^2.
\eea
\label{eq5}
\eesub
The relevant parameters as well as physical masses of different fields are listed in table \ref{tab:paramII}. The other parameters which do not find mention in this table can be chosen as per phenomenological requirements and we do not list them in this table as they do not affect our numerical calculations.

In order to compute the abundance of $N_1$, we first write down two Boltzmann equations corresponding to the evolution of $\Phi^{\pm}_2$ as well as $N_1$. They can be written in terms of their comoving number densities $(Y=n/s, n \equiv {\rm number \; density}, s \equiv {\rm entropy \;density})$ as,

\begin{equation}
\dfrac{dY_{\Phi^{\pm}_2}}{dz}=-\dfrac{\langle\sigma v\rangle\,s}{H\,z}\left(Y_{\Phi^{\pm}_2}^2-(Y^{\rm eq}_{\Phi^{\pm}_2})^2\right)-\frac{\langle \Gamma_{\Phi^{\pm}_2 }\rangle}{H \,z} Y_{\Phi^{\pm}_2}
\end{equation}
\begin{equation}
\dfrac{dY_{N_1}}{dz}=\frac{\langle \Gamma_{\Phi^{\pm}_2 }\rangle}{H \,z} Y_{\Phi^{\pm}_2},
\label{eq:N}
\end{equation}
with $z=M_{\Phi^{\pm}_2}/T$ and thermally averaged decay width of  $\Phi^{\pm}_2$
\begin{align}
 \langle \Gamma_{\Phi^{\pm}_2} \rangle &=\frac{Y^2_3}{8\pi} M_{\Phi^{\pm}_2}\left(1-\left(\frac{M_{N_1}+m_\tau}{M_{\Phi^{\pm}_2}}\right)^2\right)^{3/2}\left(1-\left(\frac{M_{N_1}-m_\tau}{M_{\Phi^{\pm}_2}}\right)^2\right)^{1/2} \frac{K_1(M_{\Phi^{\pm}_2}/T)}{K_2(M_{\Phi^{\pm}_2/}T)} \nonumber \\
  &= \Gamma_{\Phi^{\pm}_2} \frac{K_1(M_{\Phi^{\pm}_2}/T)}{K_2(M_{\Phi^{\pm}_2/}T)},
 \label{eq:decay_phi}
\end{align}
where $K_i$'s are the modified Bessel functions. Now, for the most part of the parameter space of the Yukawa $Y_3$ the freeze-in process happens while $\Phi^{\pm}_2$ is still in equilibrium. So, we can safely integrate the equation \eqref{eq:N}. So, the solution of the equation \eqref{eq:N} is given as 
\begin{table}[h]
\centering
\begin{tabular}{|c|c|}
\hline
 Parameters & Benchmark \\
\hline
$\mu_2$ & 1 TeV \\
$\mu_{3}$ & 6 TeV \\
$\mu_{4}$ & 6 TeV \\
$Y_3$ & $2.92\times 10^{-10}$ \\
$Y_2,~Y_4,~Y_5$ & 3.5 \\
$M^{\pm}_{\Phi_2}$ & 1 TeV\\
$M^{I}_{\Phi_2}$ & 1.045 TeV\\
$M^{R}_{\Phi_2}$ & 1.073 TeV\\
$\lambda_1$ & 0.255 \\
$\lambda_2$ & 0.1  \\
$\lambda_3$ & $0$ \\
$\lambda_4$ & $4.0$ \\
$\lambda_5$ & $1.0$ \\
$M_{N_1}$ & 30 MeV \\
$M_E$ & 7 TeV  \\
\hline
\end{tabular}
\caption{\label{tab:paramII} Numerical values of different relevant parameters used for Model-I.}
\end{table}

\begin{figure}[t]
\centering
\includegraphics[scale=0.5]{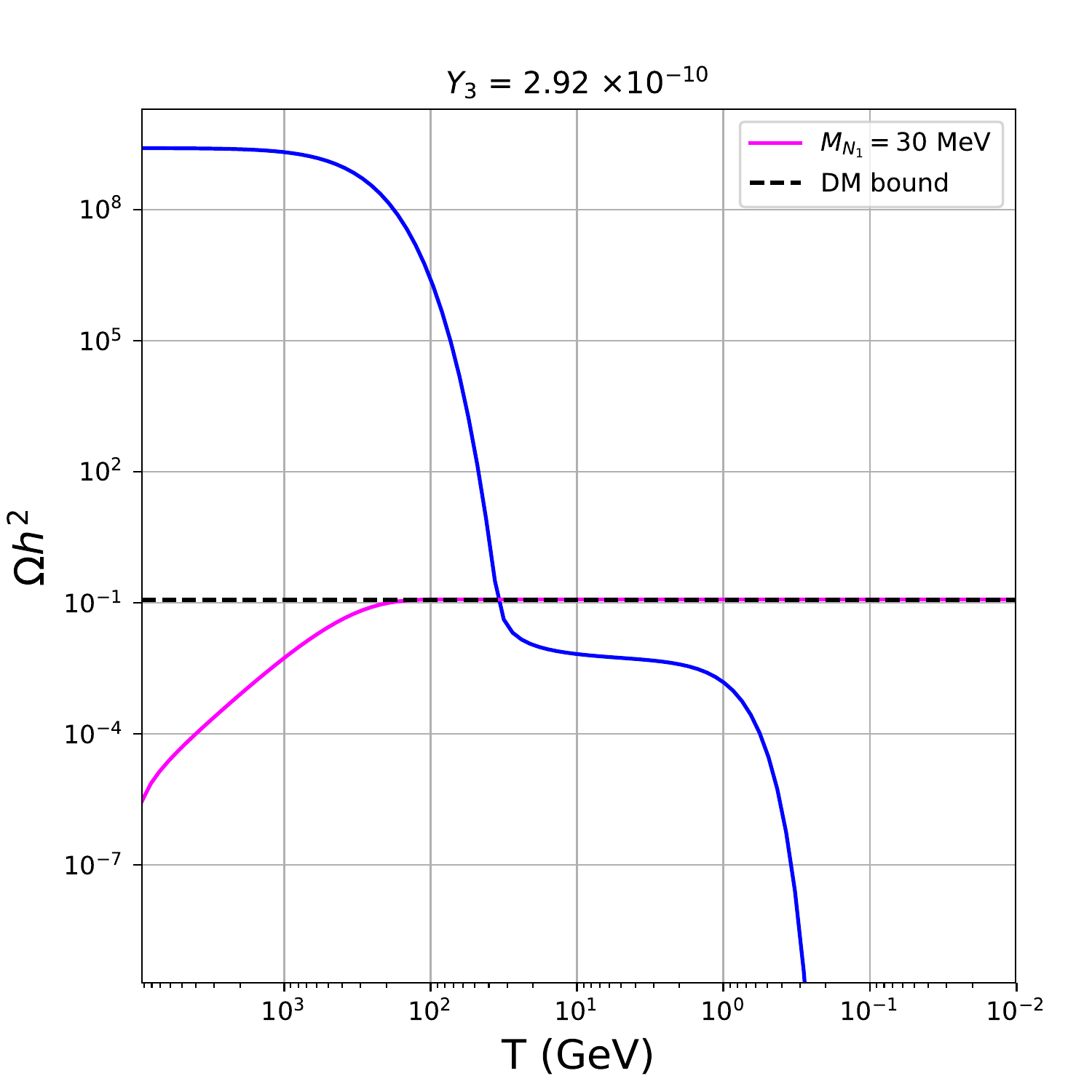}
\caption{Evolution of comoving number densities for non-thermal DM $(N_1)$ (solid magneta line) and lightest component of $\Phi_2$ (solid blue line). The dashed horizontal line corresponds to Planck limit on DM relic.}
\label{Fig:model2}
\end{figure}

\begin{figure}
\centering
\includegraphics[scale=0.5]{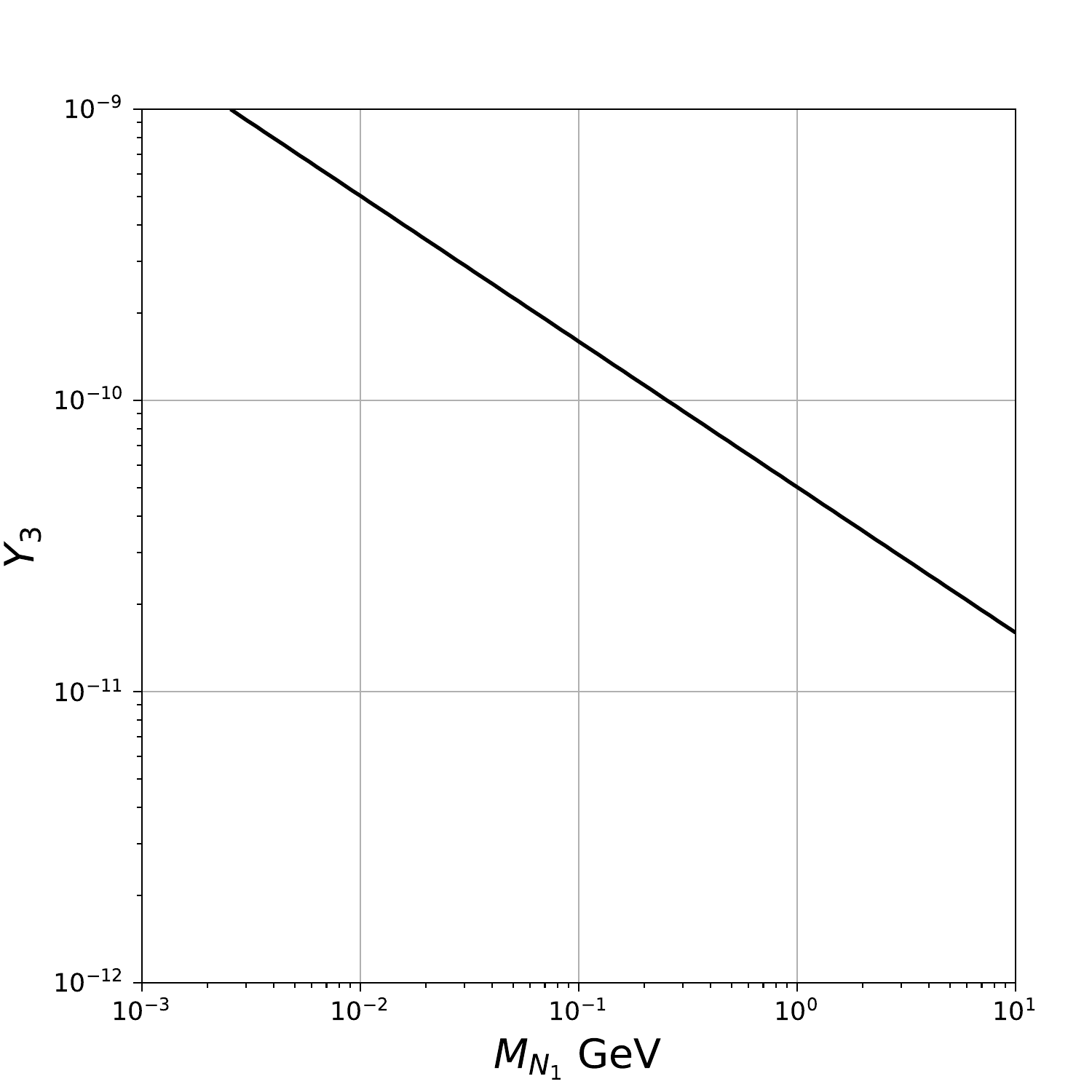}
\caption{Parameter space in $Y_3-M_{N_1}$ plane which give rise to correct DM relic.}
\label{fig:scan}
\end{figure}

\begin{align}
Y_{N_1} &= \int^\infty_{0.1}z^3K_1(z)dz\left[\frac{\Gamma_{\Phi^{\pm}_2}}{16\pi^4 H(M_{\Phi^{\pm}_2})} \frac{45}{g_*}\right] \nonumber \\
&\simeq 4.71239  \left[\frac{\Gamma_{\Phi^{\pm}_2}}{16\pi^4 H(M_{\Phi^{\pm}_2})} \frac{45}{g_*}\right]
\end{align}
and the relic is given as 
\begin{align}
\Omega_{DM} h^2 &= M_{N_1} \frac{\rho_0}{\rho_c}Y_{N_1} \simeq 2.345 \times 10^{25}\left(\frac{Y_3^2 M_{N_1}}{g_*(M_{\Phi^{\pm}_2})M_{\Phi^{\pm}_2}}\right)
\end{align}

For a chosen benchmark point, the temperature evolution of comoving number densities for DM and its mother particle are shown as function of temperature in figure \ref{Fig:model2}. We choose the benchmark point as $M_{N_1}=30 \; {\rm MeV}, M_{\Phi^{\pm}_2} = 1 \; {\rm TeV}, Y_3 = 2.92 \times 10^{-10}$ and find that the correct relic density is satisfied. It can be clearly seen from figure \ref{Fig:model2} that the number density of DM that is $N_1$ increases as temperature cools down due to decay of different components of $\Phi_2$. The equilibrium contribution of $\Phi^{\pm}_2$ decay dominates over its post freeze-out contribution to the abundance of $N_1$ for the chosen benchmark. While $N_1$ receives non-thermal contribution from all the components of $\Phi_2$, the coannihilations among different components of $\Phi_2$ have been incorporated following Ref. \cite{Griest:1990kh}. While the plot shown in figure \ref{Fig:model2} corresponds to a particular benchmark point of DM mass and Yukawa coupling, we show the allowed parameter space in terms of these two parameters in figure \ref{fig:scan} which gives rise to correct DM relic. While $Y_3, M_{N_1}$ are being varied to generate this plot, the other parameters are kept fixed at values mentioned in table \ref{tab:paramII}. The observed correlation between $Y_3$ and $M_{N_1}$ from this plot agrees with the approximate analytical expression for DM relic density derived above. Since freeze-in abundance increases with increase in DM coupling \cite{Hall:2009bx}, the corresponding DM mass needs to be smaller in order to be in agreement with observed abundance. It should also be noted that the requirement of such small Yukawa coupling of DM $N_1$ for its non-thermal nature effectively decouples one of the singlet neutrinos from the light neutrino mass generation mechanism at one loop level. The other two heavy singlet neutrinos $N_{2,3}$ can however, have order one couplings with leptons and can generate the required light neutrino mass. Therefore, the model predicts vanishingly small lightest neutrino mass. Although DM relic is satisfied for this model, but it can not explain the AAEs as we discuss in the upcoming section. Before that, we outline Model-II which satisfies all relevant DM constraints apart from explaining the AAEs.

\subsection{Model-II: Freeze-out + Freeze-in}
\label{sbsc:model2}
In this subsection, we discuss our Model-II where DM relic is generated from a hybrid setup consisting of both freeze-out and freeze-in contributions. The particle content of this model is shown in table \ref{tab:a}. There exist six different types of fields, apart from the SM fields, all of which are odd under an in-built $Z_2$ symmetry.  The SM fields, on the other hand, are even under $Z_2$ symmetry. In order to generate correct neutrino mass, we require at least two copies of right handed singlet neutrinos $N$. While the minimal scotogenic model has only two different types of $Z_2$ odd particles, the roles of other four types of $Z_2$ odd particles in this model will become clear when we discuss the explanation of the AAEs in upcoming section.

\begin{table}[h]
\centering
\begin{tabular}{|c|cccccc|cccccc|}
\hline
Fields & $Q$ & $u_R$ & $d_R$ & $L$  & $e_{R}$  &  $\Phi_1$ & $\Phi_2$ & $\Phi_3$ & $E_{L,R}$ & $N$& $U_{L,R}$ & $D_{L,R}$  \\
\hline
$SU(3)$ & 3 & 3 & 3 & 1 & 1 & 1 & 1 &1 &1 & 1 & 3 & 3 \\
\hline
$SU(2)$ & 2 & 1 & 1 & 2 & 1 & 2 & 2 & 2 &1 & 1 & 1 & 1 \\
\hline
$U(1)_Y$ & $\frac{1}{6}$ & $\frac{2}{3}$ & $-\frac{1}{3}$ & -$\frac{1}{2}$ & -1 & $\frac{1}{2}$ & $\frac{1}{2}$ & $\frac{1}{2}$ & $\pm 1$ & 0 & $\pm \frac{2}{3}$ & $\mp \frac{1}{3}$ \\
\hline
$Z_2$ & 1 & 1 & 1 & 1 & 1 & 1 & -1 &-1 & -1 & -1 & -1 & -1 \\
\hline
\end{tabular}
\caption{\label{tab:a} Field content and transformation properties under the symmetry of the model. }
\end{table}
The relevant part of the Yukawa Lagrangian is given by,
\begin{align}
\label{eq:YukMdl1}
\mathcal{L}_Y & \supset Y_1 \overline{L} \Phi_1 e_R + Y_2 \overline{L} \Phi_2 E_R + Y_3 \overline{L} \tilde{\Phi_2} N + Y_4 \overline{Q} \tilde{\Phi_2} U_R + Y_5 \overline{Q} \Phi_2 D_R   \nonumber \\
&  + Y'_2 \overline{L} \Phi_3 E_R + \tilde{Y'_3} \overline{L} \tilde{\Phi_3} N + Y'_4 \overline{Q} \tilde{\Phi_3} U_R + Y'_5 \overline{Q} \Phi_3 D_R \nonumber \\
& +Y''_2 L^T \tilde{\Phi_2} E_L + +Y''_3 L^T \tilde{\Phi_3} E_L.
\end{align} 
The scalar potential is,
\begin{align}
V(\Phi_1,\Phi_2, \Phi_3) &=  \mu_1^2|\Phi_1|^2 +\mu_2^2|\Phi_2|^2+\mu_3^2|\Phi_3|^2+\frac{\lambda_1}{2}|\Phi_1|^4+\frac{\lambda_2}{2}|\Phi_2|^4+\frac{\lambda_3}{2}|\Phi_3|^4 \nonumber \\
& +\lambda_4|\Phi_1|^2|\Phi_2|^2 +\lambda_5|\Phi_1^\dag \Phi_2|^2 + \left\lbrace\frac{\lambda_6}{2}(\Phi_1^\dag \Phi_2)^2 + \text{h.c.}\right\rbrace \nonumber \\
& +\lambda'_4|\Phi_1|^2|\Phi_3|^2 +\lambda'_5|\Phi_1^\dag \Phi_3|^2 + \left\lbrace\frac{\lambda'_6}{2}(\Phi_1^\dag \Phi_3)^2 + \text{h.c.}\right\rbrace \nonumber \\
& + \{\mu^2_{23} \Phi^{\dagger}_2 \Phi_3 + \text{h.c}\} + \lambda_7|\Phi_2|^2|\Phi_3|^2 +\lambda_8|\Phi_2^\dag \Phi_3|^2 \nonumber \\
& +\{\lambda'_7 |\Phi_1|^2 \Phi^{\dagger}_2 \Phi_3 + \lambda'_8 \Phi^{\dagger}_2 \Phi_3 |\Phi_3 |^2 + \lambda'_9 \Phi^{\dagger}_3 \Phi_2 |\Phi_2 |^2 + \text{h.c.} \} \nonumber \\
& + \left\lbrace\frac{\lambda_9}{2}(\Phi_2^\dag \Phi_3)^2 + \text{h.c.}\right\rbrace.
\label{c}
\end{align}

We assume the neutral component $\Phi^0_3$ of $\Phi_3$ to be the lightest $Z_2$ odd particle and hence the DM candidate while $E$ is the NLSP. Due to the long lived singlet fermion $E$ which decays at late times to dark matter and a tau lepton, we can have a non thermal contribution to the freeze-out abundance of DM. The physical masses of the components of scalar doublets $\Phi_2, \Phi_3$ after EWSB can be found in a way similar to Model-I. However, due to the presence of these two $Z_2$ odd scalar doublets, there can be mixing between them as well. For simplicity, we ignore such mixing, which amounts to tuning the respective bilinear and quartic couplings like $\mu_{23}, \lambda_7, \lambda_8, \lambda_9$ in the scalar potential \eqref{c}.

The coupled Boltzmann equations can be written as,
\begin{equation}
\frac{dn_{\Phi^0_3}}{dt}+3Hn_{\Phi^0_3} = -\langle \sigma v \rangle_{ \Phi^0_3\; \Phi^0_3 \rightarrow \rm SM \; SM} (n^2_{\Phi^0_3} -(n^{\rm eq}_{\Phi^0_3})^2) +N_{\Phi^0_3} \Gamma_E n_E
\label{BEforDM}
\end{equation}
\begin{equation}
\frac{dn_{E}}{dt}+3Hn_{E} = -\langle \sigma v \rangle_{E E \rightarrow \rm SM \; SM} (n^2_{E} -(n^{\rm eq}_{E})^2)-\Gamma_E n_E.
\label{BEforE}
\end{equation}
Here, $N_{\Phi^0_3}$ is the average number of DM particles produced from a single decay of $E$ which is one in this case, $\Gamma_E$ is the decay width of $E$, and $H$ is the Hubble parameter. Assuming that the singlet fermion $E$ do not contribute dominantly to the total energy budget, we can take the comoving entropy density ($g_{\star s}$) and the comoving energy density ($g_\star$) to be approximately constant. Further, we assume that almost all of $E$ decays during the radiation dominated epoch, which is satisfied for the benchmark values of Yukawa couplings we use here. Also, if we assume $E$ decays only after its thermal freeze-out, the first term on the right hand side of equation \eqref{BEforE} can be ignored \footnote{Note that the first term on the right hand side of equation \eqref{BEforE} can also be ignored if $E$ decays while in thermal equilibrium $n_E=n^{\rm eq}_E$, we consider the general case in our numerical calculations.} and one can analytically solve the Boltzmann equation for $n_E$ above. Writing the above equations in terms of $Y_{\Phi^0_3}=\frac{n_{\Phi^0_3}}{s}, Y_{E}=\frac{n_E}{s}$ with $s = \frac{2\pi^2}{45}g_{*s} T^3$ being the entropy density and changing the variable from time $t$ to $z=M_{\Phi^0_3}/T$, where $M_{\Phi^0_3}$ is the mass of DM $\Phi^0_3$, we get,
\begin{equation}
\dfrac{dY_{\Phi^0_3}}{dz}=-\dfrac{\langle\sigma v\rangle\,s}{H\,z}\left(Y_{\Phi^0_3}^2-(Y^{\rm eq}_{\Phi^0_3})^2\right)+\frac{N_{\Phi^0_3} \Gamma_N}{H \,z} Y_E
\end{equation}
\begin{equation}
\dfrac{dY_E}{dz}= -\frac{\Gamma_E}{H\,z} Y_E.
\end{equation}
The equation for $Y_E$ can be solved analytically giving,
\begin{equation}
Y_E(z) = Y_E (z_F) \text{exp} \left(-\dfrac{r}{2}\left(z^2-z_F^2\right)\right).
\end{equation}
Here, $z_F=M_{\Phi^0_3} / T_F$ is the point of freeze-out and usually takes a value of $\mathcal{O}(20)$. Also, $r=\dfrac{\Gamma_E}{H\,z^2}=\dfrac{\Gamma_E\,M_{\rm Pl}}{\pi M_{\Phi^0_3}^2}\sqrt{90/g_{\star}}$, depends on the decay width of the mother particle $E$. $Y_E(z_0)$ (where $z_0$ corresponds to the epoch after the freeze-out of $E$ from where we start integrating the differential equation for DM) however, depends on the initial abundance of $E$ and can be found by calculating its freeze-out abundance. Using this solution for $Y_E$, the equation for $Y_{\Phi^0_3}$ can be rewritten as,
\begin{equation}
\dfrac{dY_{\Phi^0_3}}{dz}=-\dfrac{\langle\sigma v\rangle\,s}{H\,z}\left(Y_{\Phi^0_3}^2-(Y^{\rm eq}_{\Phi^0_3})^2\right)+N_{\Phi^0_3}\,r\,z\,Y_E(z_0)\text{exp}\left(-\dfrac{r}{2}\left(z^2-z_0^2\right)\right).
\label{y_eq}
\end{equation}
The decay width of fermion $E$ (assuming $Y'_3=Y^{''}_3 = Y'_2/2$), ignoring the mass of leptons is given by,
\begin{equation}
\label{eq:decay_E}
\Gamma_E = \dfrac{Y^{\prime 2}_2}{16\pi}M_{E}\left(1-\dfrac{M^2_{\Phi^0_3}}{M^2_{E}}\right)^2
\end{equation}
where $M_E$ is the mass of fermion $E$ and $Y^\prime_2$, its coupling with the tau lepton and DM. The lifetime requirement of $E$ from ANITA anomaly point of view is~\cite{Fox:2018syq},
\begin{align}
\tau_E = \frac{1}{\Gamma_E} = 10 \left( \frac{M_E}{500 \; \rm GeV} \right) \; \rm ns.
\end{align}
For $M_E \sim 5$ TeV, the required lifetime of 100 ns, can be obtained for $Y^{\prime}_2 \approx 10^{-10}-10^{-9}$, typical coupling for non-thermal dark matter. It should be noted that $E$ is produced by interactions of high energy neutrinos with nucleons in a process mediated by the charged component of $\Phi_2$. The production cross section of $E$ in this process can be large by suitable tuning of the corresponding Yukawa couplings $Y_2, Y_4, Y_5, Y^{''}_2$. By keeping $\Phi_2$ heavier than $E$, in order to forbid the latter's decay into the former, we can have long lived $E$ which decays after traveling through earth into dark matter and tau lepton. We will discuss the details of AAEs in this model in the upcoming section.


\begin{table}[h]
\centering
\begin{tabular}{|c|c|c|}
\hline
 Parameters & First benchmark & Second benchmark\\
\hline
$\mu_2$  & 400 GeV & 300 GeV \\
$Y^\prime_2, ~Y^{''}_2$  & $1.1\times10^{-9}$ & $4.9\times 10^{-10}$\\
$Y_2,~Y_4,~Y_5$ & 0.96 & 1.27  \\
$\lambda_1$ & 0.255 & 0.255 \\
$\lambda_2$ & 0.1 & 0.1 \\
$\lambda_3$ & 0.81 & 0.81 \\
$\lambda_4$ & -0.3 & -0.3 \\
$\lambda_5$ & $1\times 10^{-8}$ & $1\times 10^{-8}$\\
$M_{\Phi^{\pm}_2}$ & 1.1 TeV & 1.1 TeV\\
$M_{\Phi^0_3}$  &  416.2 GeV & 321.4 GeV\\
$M_{\Phi^\pm_3}$  &  427.6 GeV & 335.9 GeV\\
$M_U$ & 5 TeV & 5 TeV\\
$M_D$ & 5 TeV & 5 TeV \\
$M_E$ & 1 TeV & 1 TeV\\
$M_N$ & 5 TeV & 5 TeV\\
\hline
\end{tabular}
\caption{\label{tab:paramI} Numerical values of different relevant parameters used for Model-II.}
\end{table}

\begin{figure}[!htbp]
\centering
\includegraphics[scale=0.45]{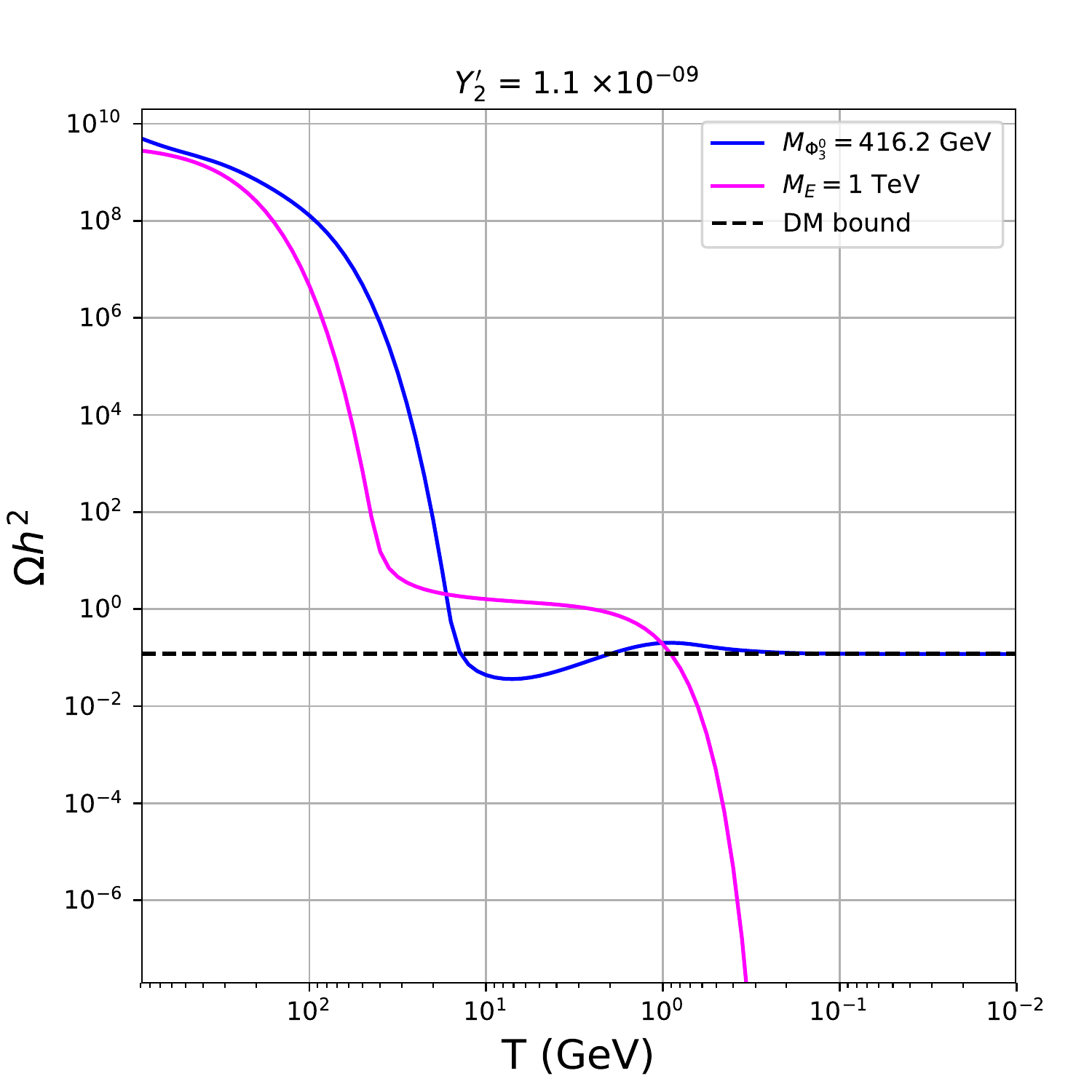}~~
\includegraphics[scale=0.45]{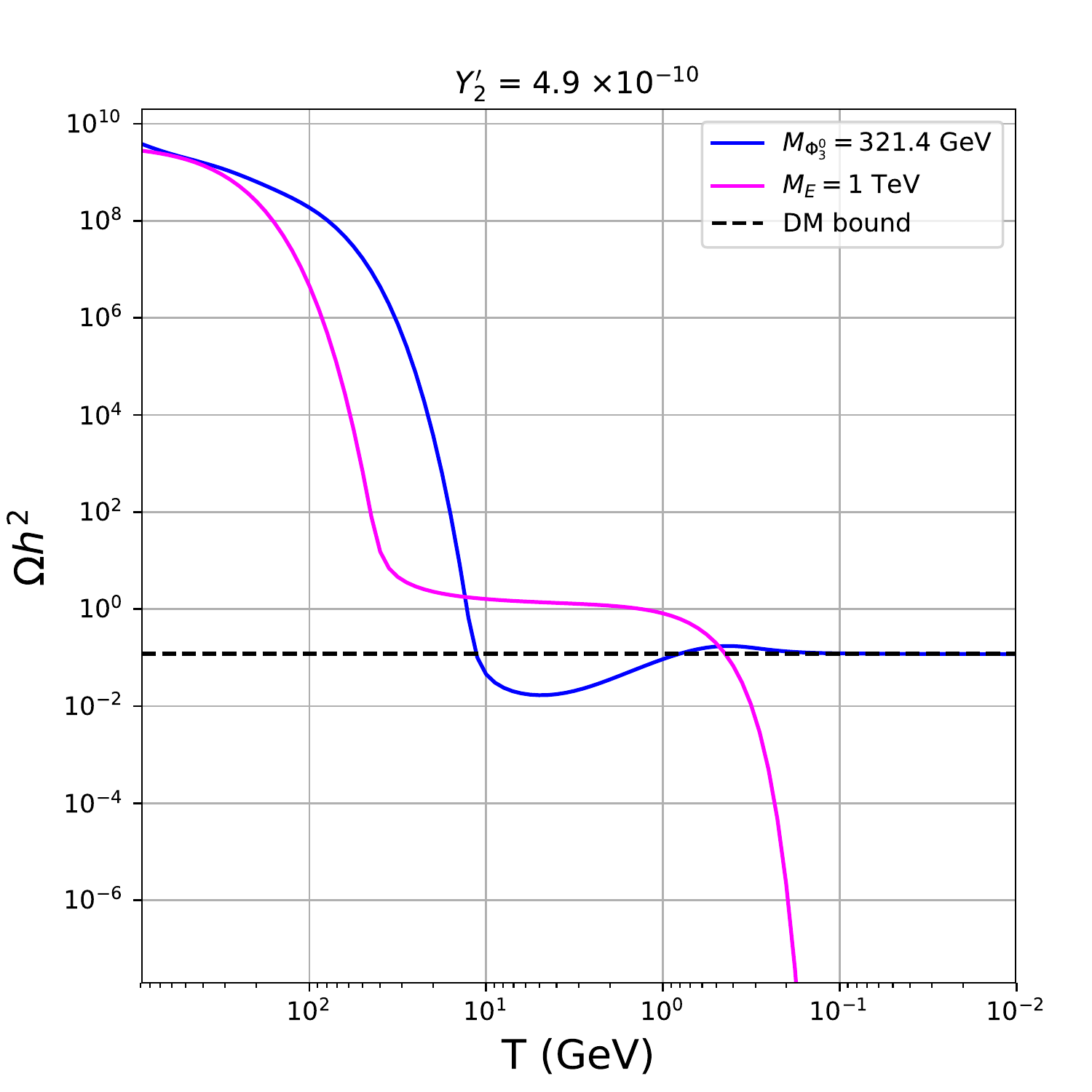}
\caption{Evolution of comoving number densities for DM $(\Phi^0_3)$ (solid blue line) and mother particle $E$ (solid magneta line) giving a non-thermal contribution to DM relic at late epochs. The dashed horizontal line corresponds to Planck limit on DM relic. The left and right panel plots correspond to two different benchmark choices for Model-II given in table \ref{tab:paramI}.}
\label{Fig:model1}
\end{figure}

In order to show the DM relic density results, we choose two benchmark points, the details of which are given in table \ref{tab:paramI}. For the chosen benchmark points, we then solve the coupled Boltzmann equations for $\Phi^0_3, E$ and show their comoving number densities as function of temperature in figure \ref{Fig:model1}. As can be seen from the plots, the comoving number densities of both $\Phi^0_3$ and $E$ decreases as temperature falls below their respective masses, owing to the Boltzmann suppression for non-relativistic species. At some epochs, the individual rate of interactions of $E$ as well as $\Phi^0_3$ with the SM particles freezes out giving rise to the plateau shaped regions. However, since $E$ is unstable and decays into $\Phi^0_3$, the comoving number density of $E$ drops very fast at some epoch due to its decay with a corresponding rise in the number density of $\Phi^0_3$. In the absence of $E$ decay, the freeze-out abundance of $\Phi^0_3$ remains under-abundant compared to the Planck bound. The mass of $\Phi^0_3$ in both the benchmark points given in table \ref{tab:paramI} falls in the mass regime $80 \; \text{GeV} \leq M_{\Phi^0_3} \leq 500 \; \text{GeV}$ where scalar doublet DM fails to give rise to correct thermal relic as is well known from earlier studies on $Z_2$ odd scalar doublet dark matter. The role of non-thermal contribution in filling this deficit was discussed in earlier work \cite{Borah:2017dfn}.



\section{Explanation of ANITA Events}
\label{sec:anitaEvts}
The estimation of number of AAEs can be facilitated from the evaluation of survival probability $\epsilon$ which basically represents the fraction of incoming ultra-high energy neutrino flux that ultimately gives rise to $\tau$-leptons near the earth surface. The total number of events can be estimated as,
\begin{align}
\label{eq:nEvts}
\mathcal{N} = A_{\rm eff} T \Delta\Omega \int_{E_{i}}^{E_{f}}dE_{\nu}\Phi(E_{\nu})\epsilon(E_{\nu}),
\end{align}
where the effective area $A_{\rm eff} \approx 4$ km$^{2}$, $T$ is the exposure time which we take $\approx25$ days by combining the exposures of ANITA-I and ANITA-III, $\Delta \Omega$ is the acceptance angle which corresponding to isotropic and anisotropic neutrino flux source is given as,
\begin{align}
\Delta \Omega \approx \begin{cases}
2\pi ~\text{sr for isotropic case,}\\
2\pi(1-\cos \delta_{\theta}) \approx 0.0022 ~\text{sr for anisotropic case,}
\end{cases}
\end{align}
where $\delta_{\theta} \approx 1.5^{\circ}$ is the angular uncertainty relative to parent neutrino direction. Note that we have not considered the exposure time for ANITA-II as it was not sensitive to such events. The range of integration in equation~\eqref{eq:nEvts} should be such that the correct range of shower energy can be obtained. As we consider $\tau$ lepton as the origin of observed events, with $E_\tau=E_\nu/4$ and observed shower energy $0.2$-1 EeV, we take $E_i=0.8$ EeV, $E_f=4$ EeV. As far as the flux $\Phi(E_{\nu})$ is concerned, for the isotropic case if one assumes the source of the EeV neutrinos to be the Greisen-Zatspin-Kuzmin (GZK) mechanism then in that case, following~\cite{Aartsen:2018vtx}, for the concerned shower energy, $\Phi_{\rm iso}\approx 10^{-25}$ (GeV cm$^{2}$ s sr)$^{-1}$. Otherwise, if some localized source is the origin of the EeV neutrinos, then the upper limit on such anisotropic flux would be $\Phi_{\rm aniso}\approx 3.2\times 10^{-20}$ (GeV cm$^{2}$ s sr)$^{-1}$~\cite{Jacobsen:2015mga, Adrian-Martinez:2015ver, Aartsen:2016oji, Mertsch:2016hcd, Fujii:2018yhp}. Thus by considering the mentioned parameters in the isotropic and anisotropic cases to get two AAEs, one should have,
\begin{align}
\mathcal{N} \approx \begin{cases}
2.0\times 10^{2}\epsilon ~\text{for isotropic case,}\\
2.0\times 10^{4}\epsilon~\text{for anisotropic case.}
\end{cases}
\label{eq:nevent}
\end{align}
This implies that for the observed two AAEs in the isotropic case the $\epsilon$ should be $\sim 10^{-2}$, and for the anisotropic case $\epsilon$ should be $\sim 10^{-4}$. By considering SM interactions, authors of~\cite{Fox:2018syq} predicted the survival probability $\epsilon_{\rm SM}\sim 10^{-7}$, giving $\mathcal{N} \sim 2\times 10^{-5}$ for the isotropic flux. Clearly in the SM explanation of these two events is very unlikely. 
While using the SM interaction and considering the anisotropic flux one gets $\mathcal{N} \sim 2 \times 10^{-3}$ which again makes it very implausible to get explanation of the observed events in the SM.
Thus in any BSM scenario trying to explain the two AAEs, one should strive to increase the $\epsilon(E_{\nu})$ and we will show that this can be achieved in one of the SM extensions that we are considering. 
\\
In the following section we discuss the shower events, generated from $\tau$ decay, which is interpreted as the origin of the AAEs. Basically, $\tau$ which induces the air shower above Antarctic surface could also pass through the IceCube detector giving similar neutrino events, but no such event is observed by IceCube collaboration. In Ref.~\cite{Fox:2018syq}, the authors have interpreted 3 events of PeV energies in IceCube data~\cite{Aartsen:2018vtx} as sub-EeV earth emergent cosmic rays similar to ANITA. The exposure of IceCube is estimated at $54.0~\rm km^2~sr~yr$ which is 20 times that of ANITA exposure, $2.7~\rm km^2~sr~yr$~\cite{Fox:2018syq}. By using the relative exposures, the number of events $\mathcal{N}$ at ANITA can be estimated as $\mathcal{N}\approx \mathcal{N_{IC}}/20$ giving $\mathcal{N}=0.15$ for $\mathcal{N_{IC}}=3$ events at IceCube which is roughly an order of magnitude away from the observed events by ANITA. As pointed out in ~\cite{Fox:2018syq}, ANITA estimated event rate serves as an upper limit which can be reduced for a better angular coverage and a detailed analysis including particular instrumental effects is needed to resolve it. In the following section, while remaining consistent with IceCube, we will estimate the required survival probability for $\mathcal{N}=0.15$. In the estimation of survival probability we have closely followed the approach of~\cite{Chauhan:2018lnq}.

As discussed in the considered scenarios, the NLSP decays to LSP and a $\tau$-lepton which is assumed to be the origin of the observed AAEs. While propagating through the earth, the NLSP may loose its energy through the electromagnetic processes. The average energy loss of a particle traveling a distance $l$ can be estimated from the equation, $-dE/dl=\beta(E) E \rho(l)$ where $\rho(l)$ is the density of the earth, and $\beta$ is the radiative energy loss. While using the parametrization for $\beta(E)$ from~\cite{Reno2005} we found $\beta=6.5\times 10^{-10}$ cm$^2/$gm for NLSP initial energy $E=1$ EeV and mass $M_{NLSP}=1$ TeV. Then with $\rho=4~\rm {gm/cm^3}$~\cite{Gandhi1996} for $l\sim 7000$ km one can find the energy loss due to radiation to be $\sim 17\%$, implying the NLSP energy $E\sim 0.8$ EeV. The corresponding final shower energy $E/2 \sim 0.4$ EeV is consistent with the ANITA observations.

\subsection{Model-I}
\label{sbsc:model2ANITA}
The basic idea is to consider the ultra-high energy (UHE) neutrino interactions with the matter inside earth. At the elementary level the incoming UHE neutrino with energy $E_{\nu}$ interacts with the quarks mediated by the appropriate new physics particle and produce the BSM particle which travels through the earth and decays to $\tau$-lepton which actually gives rise to the AAEs. Given the particle content shown in table \ref{tab:a2} and interactions in equation \eqref{yukawamodel1} the relevant processes in this case are $\nu \bar{d} \to \Phi_{2}^{-}\psi_{2}$ and $\nu \bar{u} \to \Phi_{2}^{-}\psi_{1}$ which are mediated by the particle $E$, where $\Phi_{2}^{-}$ is the charged scalar arising from the doublet $\Phi_{2}$. In this case the final $\tau$-lepton is obtained from the decay of the $\Phi_{2}^{-}$, which decays to $N_1$ and $\tau$. In this model, $N_1$ is identified as DM candidate.
\begin{figure}[!htbp]
\centering
\includegraphics[scale=0.5]{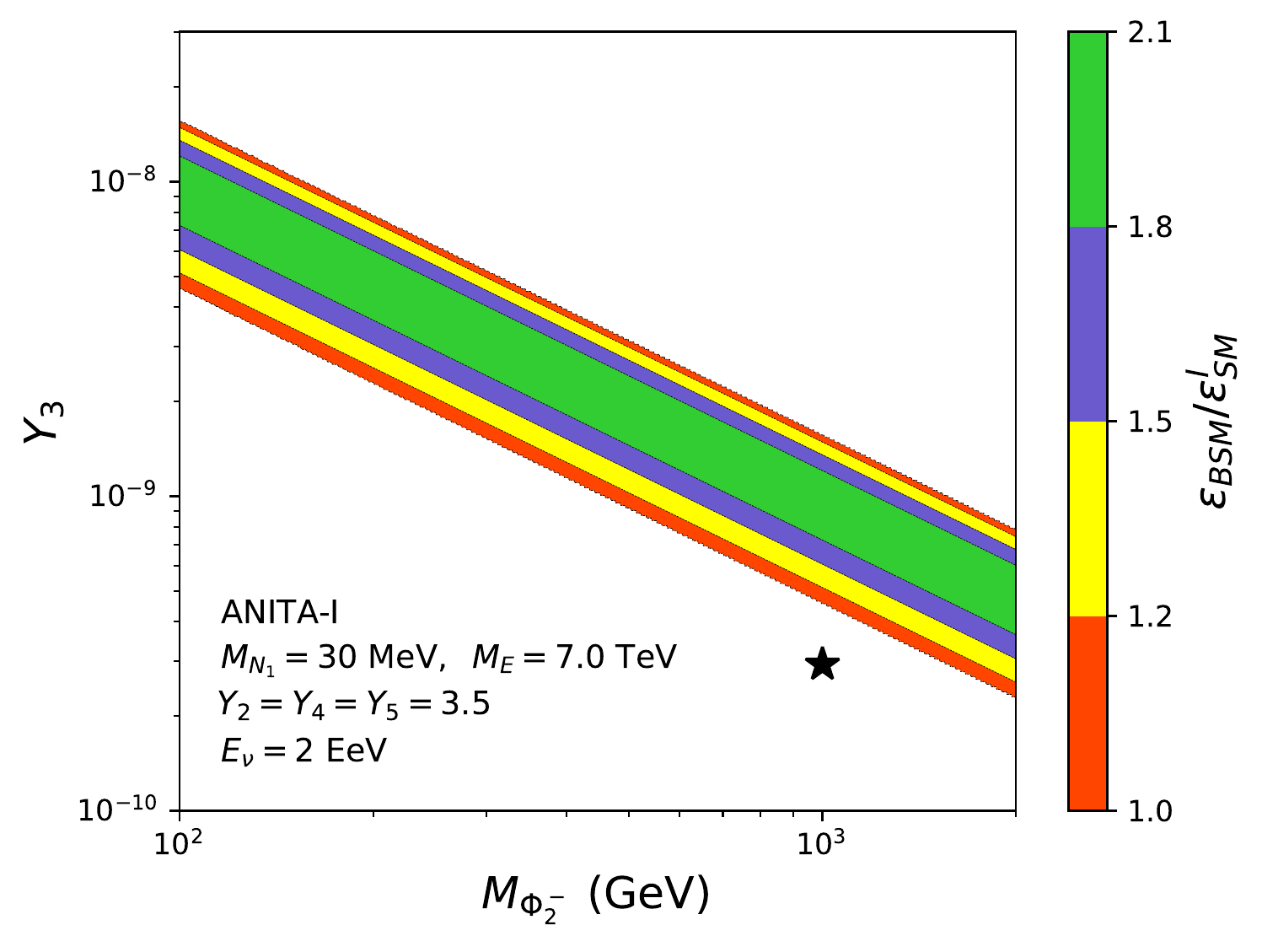}~~
\includegraphics[scale=0.5]{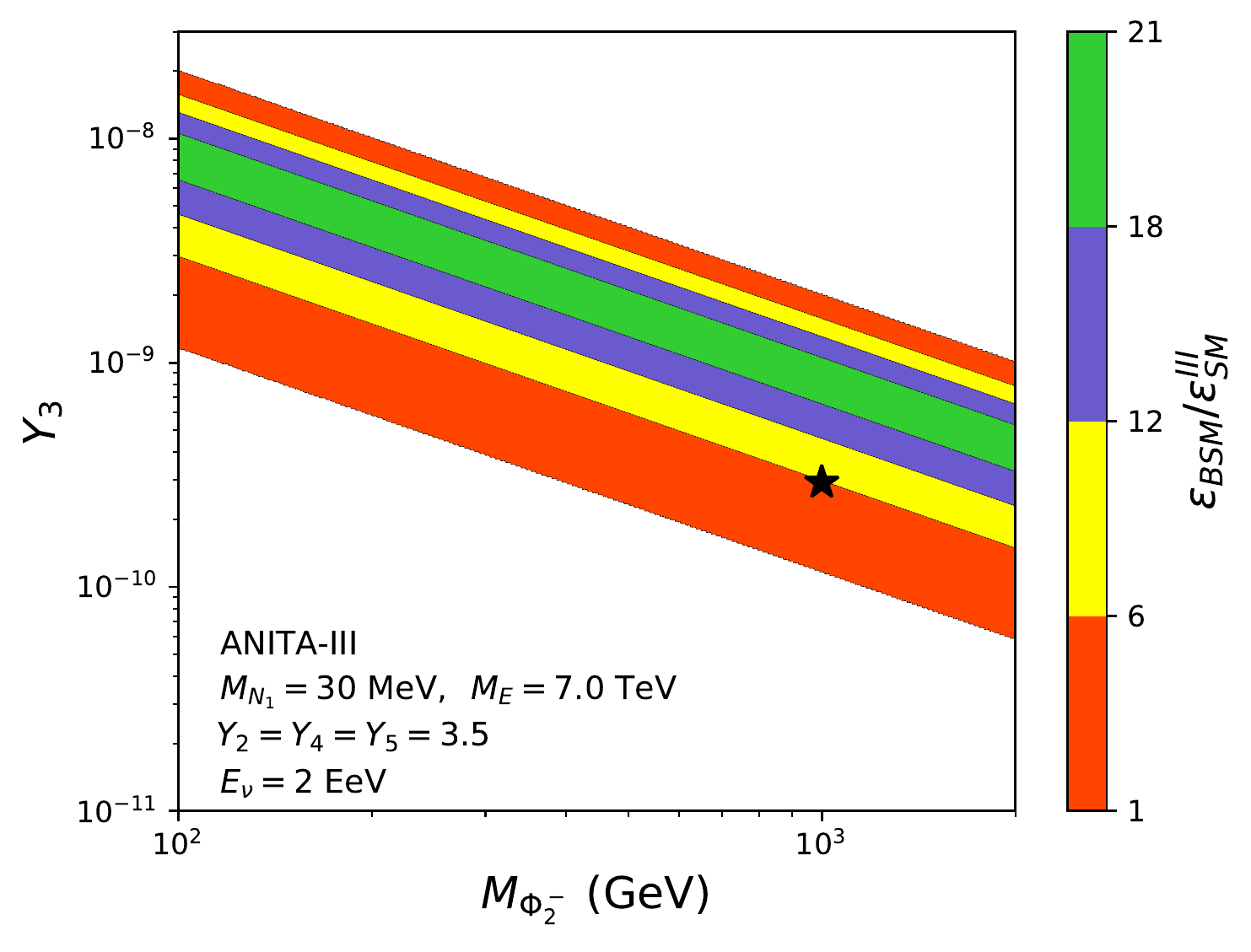}
\caption{Model-I: The survival probability $\epsilon_{BSM}$ of $\Phi^-_2$ normalized to $\epsilon^I_{SM}$ (left) and $\epsilon^{III}_{SM}$ (right) as a function of $M_{\Phi^-_2}$ and its decay coupling $Y_3$ to DM and tau lepton. Here $\epsilon^I_{SM}$ and $\epsilon^{III}_{SM}$ are SM survival probabilities for  ANITA-I and ANITA-III events respectively. The benchmark consistent with the relic density calculation is shown as a star.}
\label{fig:model2a}
\end{figure}
In terms of Bjorken scaling variables $x=Q^2/2m_N E^\prime_{\nu}$, $y=E^\prime_\nu/E_\nu$, where $m_N=(m_p+m_n)/2$, the average mass of proton and neutron, $E^\prime_\nu=E_\nu-E_{\Phi^-_2}$ is the energy loss in the lab frame, $-Q^2$ the invariant momentum transfer between neutrino (antineutrino) and particle $\Phi^-_2$, the differential cross-section of neutrino (antineutrino) nucleon scattering can be written as,
\begin{equation}
 \frac{d^2\sigma}{dx dy}=\frac{Y^2_2}{32\pi(Q^2+M^2_{E})^2}\times x s y(1-y)(Y^2_4+Y^2_5)(f_{\bar u}(x,Q^2)+f_{\bar d}(x,Q^2)),
 \label{eq:dcsmod2}
\end{equation}
where $s=2m_N E_\nu$ is the center of mass energy, $f_q(x,Q^2)$ and $f_{\bar q}(x,Q^2)$ are the parton distribution functions (PDFs) of quarks and antiquarks with $q=u,d$. In our analysis, we consider the standard $(1:1:1)$ flavor ratio of neutrino (antineutrino) on earth and use \texttt{CTEQ6} leading order PDF sets~\cite{Pumplin2002} while calculating the cross-sections.
By using the interaction cross-section obtained from equation~\eqref{eq:dcsmod2}, the interaction length, $l_{\Phi_2}$, can be written as, 
\begin{equation}
 l_{\Phi^-_2}=\frac{1}{\rho \sigma N_A},
 \label{eq:intlenmod2}
\end{equation}
where $N_A=6.022\times 10^{23}~\rm cm^{-3}$ is the Avogadro number and $\rho\approx 4$ is the earth density in water equivalent unit. A convenient representation of the density profile of the earth is given in~\cite{Gandhi1996} which is used to calculate the approximate value of $\rho$ for the relevant chord lengths for the AAEs.
The decay length of $\Phi^-_2$, $l_D$, in the earth rest frame can be written as,
\begin{equation}
\label{eq:lDMdl2}
 l_D=\gamma c \tau = \frac{1}{\Gamma_{\Phi^-_2}}\frac{E_{\Phi^-_2}}{M_{\Phi^-_2}},
\end{equation}
where $\Gamma_{\Phi^-_2}\equiv\Gamma({\Phi^-_2\rightarrow \tau^-N_1})$ is the rest frame decay width of $\Phi^-_2$ into $\tau$ lepton and DM $N_1$ which is obtained from equation~\eqref{eq:decay_phi}.
Here we used the approximation $E_{\Phi^-_2}=(1-\xi)E_\nu/2$, where $\xi$ accounts for the energy loss of ${\Phi^-_2}$ due to radiation (which is $\sim 17\%$ as estimated before), to get the observed shower energy $E_\tau=E_{\Phi^-_2}/2 \sim 0.4$ EeV by taking into account the incident neutrino energy $E_\nu=2$ EeV. As mentioned before, by considering only the SM interactions it is not possible to get the desired survival probability for any SM particle to pass a chord distance of $\sim 7000$ km. In considered BSM scenarios, as a result of additional interactions, the survival probability of the neutrino flux can be estimated as~\cite{Chauhan:2018lnq},
\begin{equation}
\label{eq:survProb1}
\epsilon_{BSM}= \int^{l_{\Earth}}_0 dl_1 \int^{l_{\Earth}-l_1}_{l_{\Earth}-l_1-h} dl_2 \frac{e^{-l_2/l_D}}{l_D}\frac{e^{-l_1/l_{\Phi_{2}^{-}}}}{l_{\Phi_{2}^{-}}}\left(1-\int^{l_1}_0 \frac{dl_3}{l_0}e^{-l_3/l_0}\right),
\end{equation}
where $l_{\Earth}$ is the chord length, $l_0$ is the mean interaction length which is $290$ and $265$ km for ANITA-I and ANITA-III events respectively~\cite{Fox:2018syq}, and $h\approx 10$ km is the window of $\tau$ production near the surface of earth.

%

In figure~\ref{fig:model2a}, we plotted the survival probability $\epsilon_{BSM}$ of $\Phi^-_2$ by normalizing it with respect to $\epsilon_{SM}$, as a function of $M_{\Phi^-_2}$, and it's decay coupling $Y_3$ with DM and tau lepton, for ANITA-I (left) and ANITA-III (right) events. While scanning the parameter space, we considered the benchmark relevant for the correct relic density which are quoted in table~\ref{tab:paramII} and shown as a star in figure~\ref{fig:model2a} for $M_{\Phi^-_2}=1$ TeV and $Y_3=2.92\times10^{-10}$. The color scheme corresponds to different regions of parameter space where $\epsilon_{BSM}/\epsilon^{I,III}_{SM} \geq 1$. In our analysis, we considered $\epsilon^I_{SM}=4.4\times 10^{-7}$ and $\epsilon^{III}_{SM}=3.2\times 10^{-8}$ for ANITA-I and ANITA-III events respectively~\cite{Fox:2018syq}. By using the benchmark in this model we get $\epsilon_{BSM}=2.45\times10^{-7}$ for ANITA-I event and $\epsilon_{BSM}=2.17\times10^{-7}$ for ANITA-III event. When these values are used in equation~\eqref{eq:nevent}, we found $\mathcal{N}=0.00009$ for isotropic flux and $\mathcal{N}=0.009$ for anisotropic flux. Clearly, in this model, it is not possible to get the desired number of events for perturbative Yukawa couplings while being consistent with DM relic density estimation.

\subsection{Model-II}
\label{sbsc:model1ANITA}
The basic tenet of explaining the AAEs in this model follows a similar path as the Model-I, mentioned in the previous subsection.  In the Model-II, outlined in subsection~\ref{sbsc:model1}, the necessary interactions can be obtained from the Yukawa Lagrangian defined in equation~\eqref{eq:YukMdl1}. In this case, the typical relevant processes would be $\nu d \to EU$ and $\nu \bar{u} \to E\bar{D}$ mediated by the charged scalar arising from the doublet $\Phi_{2}$. The produced $E$ subsequently decays to $\tau$ and the DM candidate which is the neutral component of $\Phi_{3}$. 
%
\begin{figure}[t!]
\centering
\includegraphics[scale=0.5]{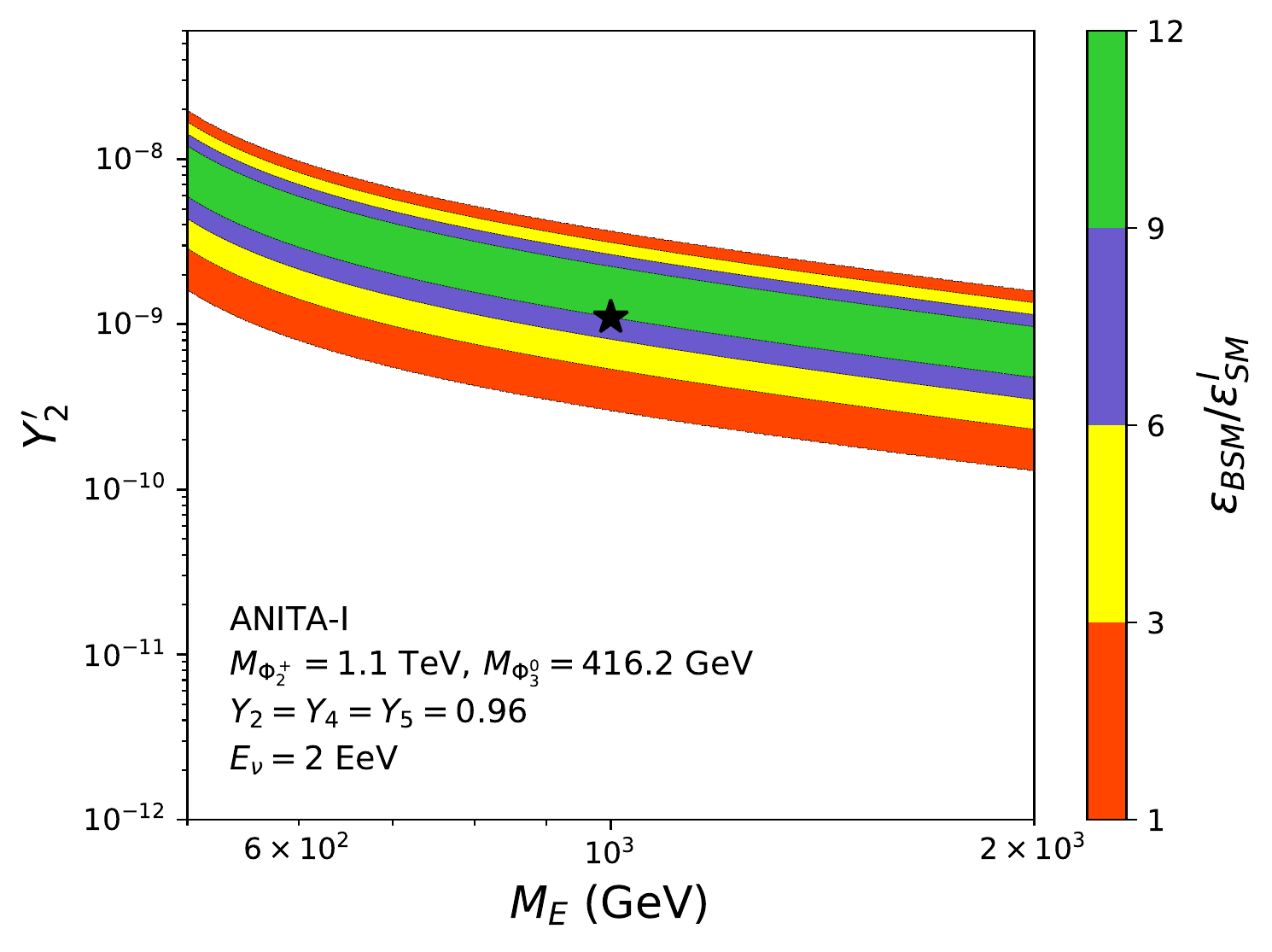}~~
\includegraphics[scale=0.5]{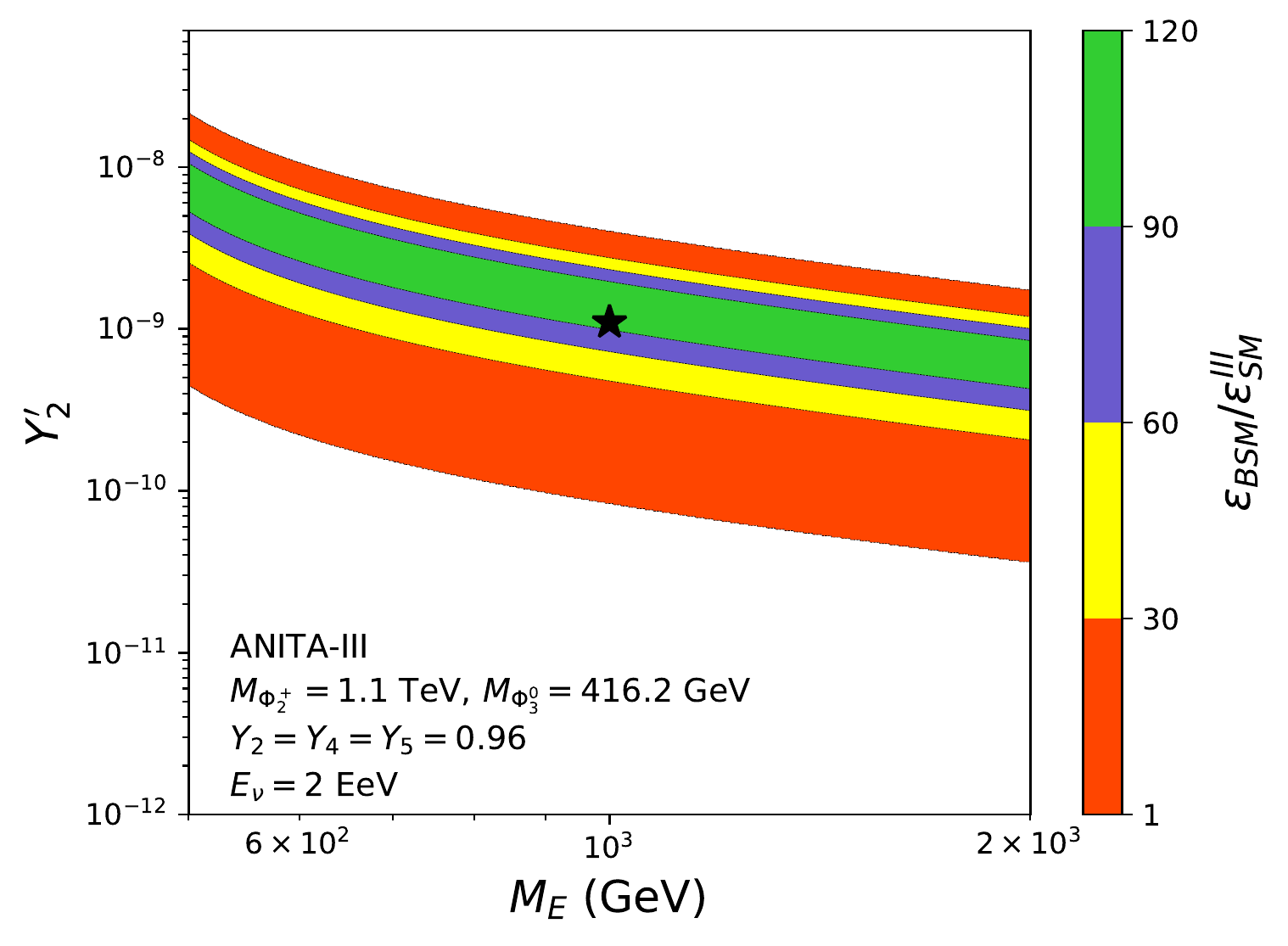}
\caption{Model-II: The survival probability $\epsilon_{BSM}$ of $E$ normalized to $\epsilon^I_{SM}$ (left) and $\epsilon^{III}_{SM}$ (right) as a function of fermion mass $M_E$ and it's decay coupling $Y^\prime_2$ to DM and tau lepton. Here $\epsilon^I_{SM}$ and $\epsilon^{III}_{SM}$ are SM survival probabilities for  ANITA-I and ANITA-III events respectively. The benchmark consistent with the relic density calculation is shown as a star. }
\label{fig:model1a}
\end{figure}
Now for the mentioned processes, the total differential cross-section of neutrino (antineutrino) scattering in terms of Bjorken scaling variables $x=Q^2/2m_N E^\prime_{\nu}$, $y=E^\prime_\nu/E_\nu$, with $E^\prime_\nu=E_\nu-E_{E}$ the energy loss in the lab frame, $-Q^2$ the invariant momentum transfer between neutrino (antineutrino) and particle $E$, can be written as,
\begin{align}
 \frac{d^2\sigma}{dx dy}=\frac{x s y^2 Y^2_2}{16\pi(Q^2+M^2_{\Phi^\pm_2})^2}\left[ Y^2_4 \left(f_u(x,Q^2)+f_{\bar u}(x,Q^2)\right) + Y^2_5 \left(f_d(x,Q^2)+f_{\bar d}(x,Q^2)\right)\right],
 \label{eq:dcsmod1}
\end{align}
where $s=2m_N E_\nu$ is the center of mass energy. Similar to the previous case the interaction length, $l_{E}$, can be given as,
\begin{equation}
 l_E=\frac{1}{\rho \sigma N_A},
 \label{eq:intlenmod1}
\end{equation}
where $\sigma$ is the interaction cross-section obtained from Eq.~\eqref{eq:dcsmod1}. 
The decay length of $E$ in the earth rest frame can be written as,
\begin{equation}
 l_D=\gamma c \tau = \frac{1}{\Gamma_E}\frac{E_E}{M_E},
\label{eq:lDMdl1} 
\end{equation}
where $\Gamma_E\equiv\Gamma({E\rightarrow \tau^-\Phi^0_3})$ is the rest frame decay width of $E$ obtained from equation~\eqref{eq:decay_E}. As before, we used the approximation $E_E=(1-\xi)E_\nu/2$, where $\xi$ accounts particle $E$ energy loss due to radiation (which is $\sim 17\%$ as estimated before), to get the observed shower energy $E_\tau=E_E/2 \sim 0.4$ EeV by taking into account the incident neutrino energy $E_\nu=2$ EeV.
%
\begin{figure}[t!]
\centering
\includegraphics[scale=0.5]{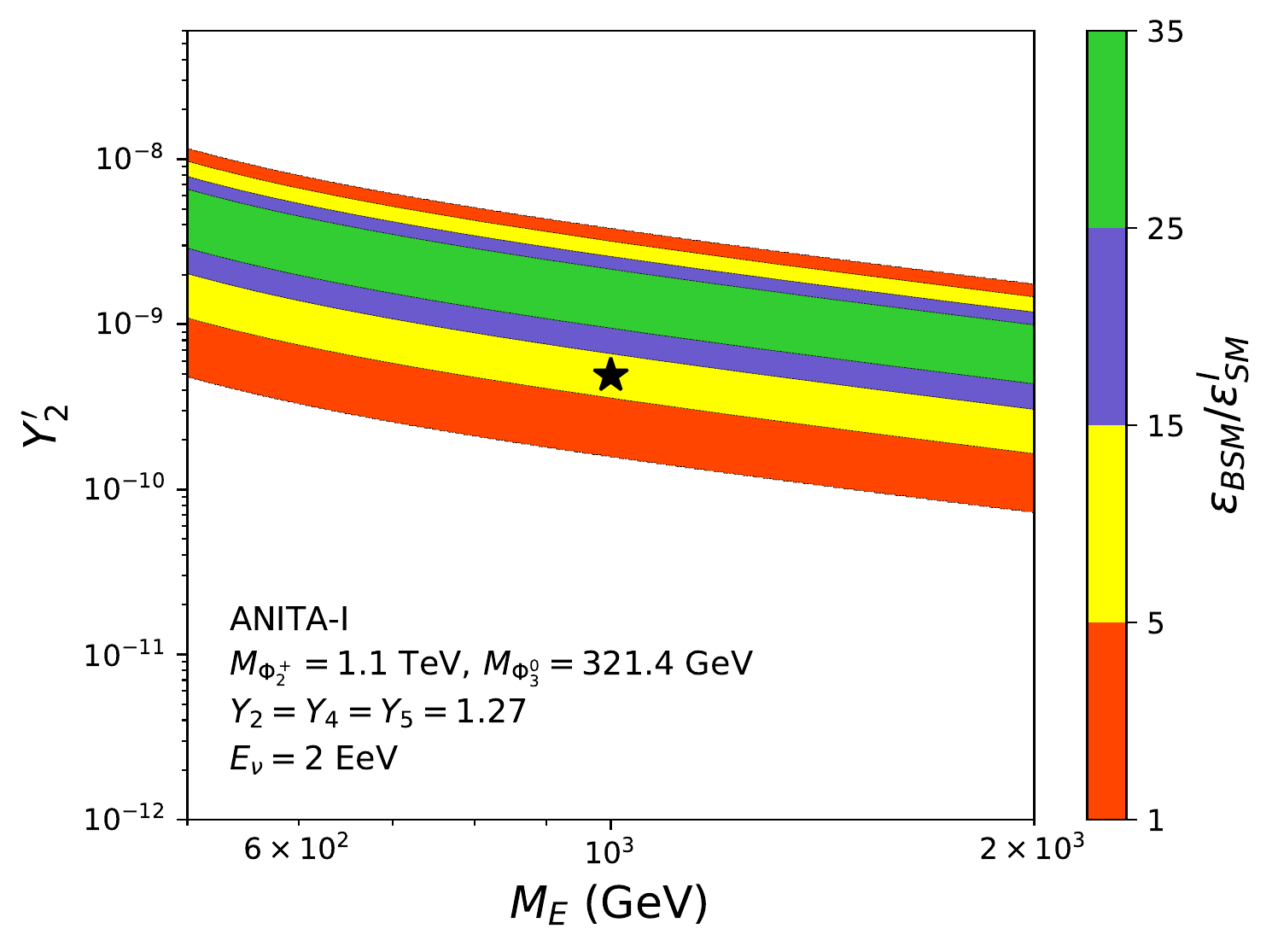}~~
\includegraphics[scale=0.5]{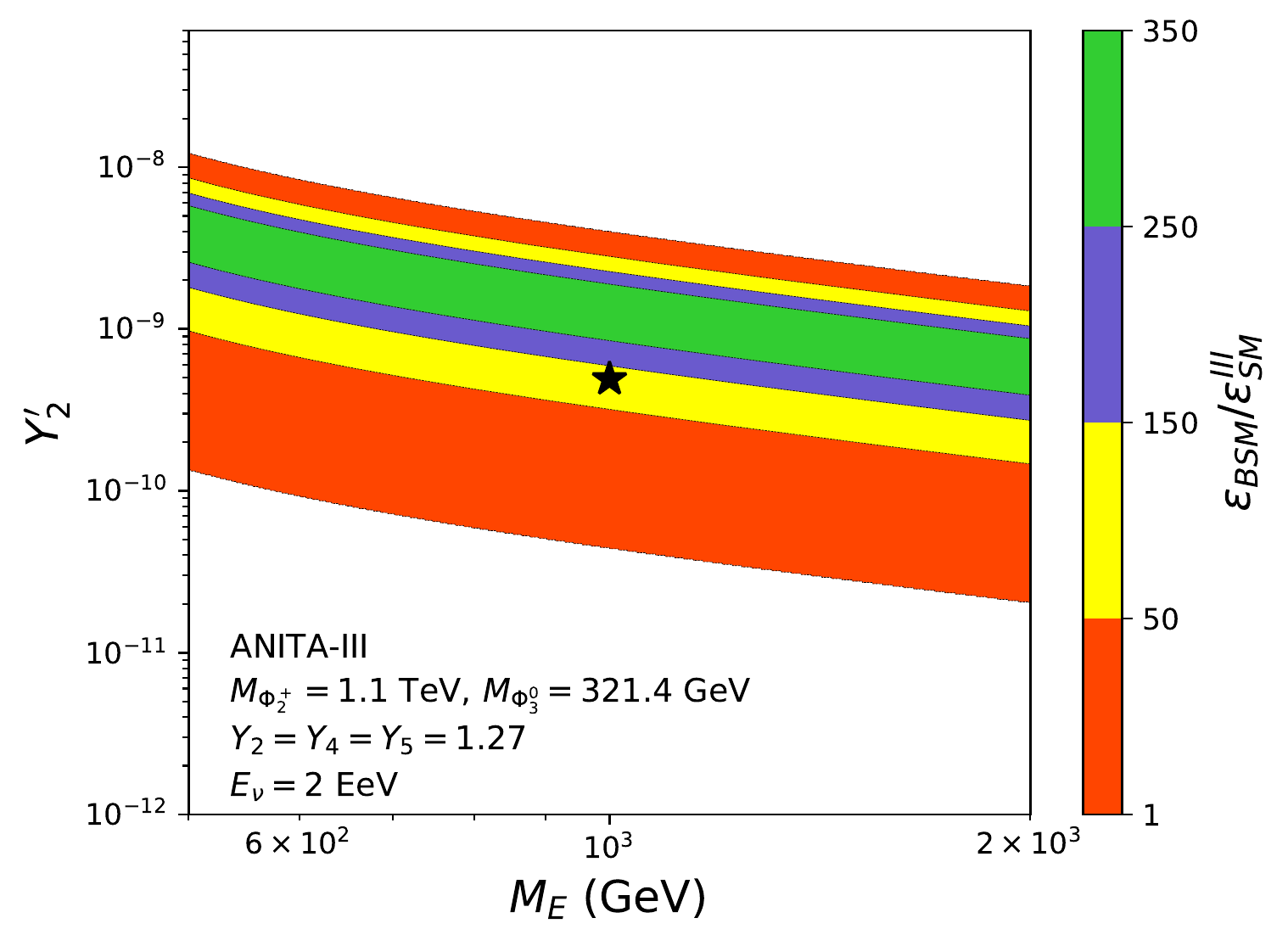}
\caption{Same as figure~\ref{fig:model1a} for the second benchmark.}
\label{fig:model1b}
\end{figure}
In this case, the survival probability of the neutrino flux, $\epsilon_{BSM}$, can also be written following equation~\eqref{eq:survProb1} by making the substitution $l_{\Phi_{2}^{-}} \to l_{E}$ and the appropriate $l_{D}$ defined in equation~\eqref{eq:lDMdl1}.
In figure~\ref{fig:model1a}, we plotted the survival probability $\epsilon_{BSM}$ of $E$ normalized to $\epsilon_{SM}$, as a function of it's mass, $M_E$, and it's decay coupling $Y^\prime_2$ to DM and tau lepton for ANITA-I (left) and ANITA-III (right) events. While scanning the parameter space, we considered the parameters relevant for correct relic density which are quoted in table~\ref{tab:paramI} and specifically first benchmark point is shown as a star in figure~\ref{fig:model1a} for $M_E=1$ TeV and $Y^\prime_2=1.1\times10^{-9}$. The color scheme corresponds to different regions of parameter space where $\epsilon_{BSM}/\epsilon^{I,III}_{SM}\geq 1$. While using the first benchmark point, we get $\epsilon_{BSM}=4.33\times10^{-6}$ for ANITA-I event and $\epsilon_{BSM}=3.41\times10^{-6}$ for ANITA-III event. These values are then used in equation~\eqref{eq:nevent} and we found $\mathcal{N}=0.0015$ for isotropic flux, and $\mathcal{N}=0.154$ for anisotropic flux.
We repeat the similar exercise for second benchmark point by taking into account the relic density estimations and get $\mathcal{N}=0.17$ for anisotropic flux.
Our result for the second benchmark point is shown in figure~\ref{fig:model1b}. Clearly for both the benchmarks, it is not possible to get the desired survival probability for isotropic flux and so anisotropic flux is required while being consistent with DM relic density estimation.



\section{Complementary Probes of the Model}
\label{sec:comp}
Apart from providing an explanation to AAEs along with dark matter, the model we propose here can also offer tantalizing signatures at different experiments operating at energy as well as intensity frontiers. We briefly comment on such possibilities related to model-II which is successful in explaining the observed AAEs discussed before. Clearly, the model is an extension of the widely studied inert Higgs doublet model or scotogenic model of radiative neutrino mass with additional $Z_2$ odd particles. The scalar sector of model-II contains two $Z_2$ odd scalar doublets $\Phi_{2,3}$. By virtue of their electroweak gauge interactions, the components of these scalar doublets can be produced significantly in proton-proton collisions of the Large Hadron Collider (LHC). Depending upon the mass spectrum of its components, the heavier ones can decay into the lighter ones and a gauge boson, which finally decays into a pair of leptons or quarks. Therefore, we can have either pure leptonic final states plus missing transverse energy (MET), hadronic final states plus MET or a mixture of both. The MET corresponds to DM or light neutrinos. In several earlier works \cite{Miao:2010rg, Gustafsson:2012aj, Datta:2016nfz}, the possibility of opposite sign dileptons plus MET was discussed. In \cite{Poulose:2016lvz}, the possibility of dijet plus MET was investigated with the finding that inert scalar masses up to 400 GeV can be probed at high luminosity LHC. In another work \cite{Hashemi:2016wup}, tri-lepton plus MET final states was also discussed whereas mono-jet signatures have been studied by the authors of \cite{Belyaev:2016lok, Belyaev:2018ext}. The enhancement in dilepton plus MET signal in the presence of additional vector like singlet charged leptons (like we have in model-II) was also discussed in \cite{Borah:2017dqx}. Exotic signatures like displaced vertex and disappearing or long-lived charged track for compressed mass spectrum of inert scalars and singlet fermion DM was studied recently by the authors of \cite{Borah:2018smz}. 
To summarise, the search strategy and the bounds on $Z_2$ odd scalars depend crucially on the model and spectrum of lighter particles to which they can decay. Due to identical gauge interactions, the bounds are somewhat similar to the ones on sleptons in supersymmetric models. As mentioned in \cite{Tanabashi:2018oca}, collider bounds on charged sleptons are also dependent upon lightest neutralino mass among other details related to the mass spectrum. Since we are not performing a detailed collider analysis in our work, we consider a conservative lower bound on such $Z_2$ odd charged scalar masses ($> 100$ GeV) in agreement with LEP bounds \cite{Lundstrom:2008ai, Tanabashi:2018oca}. Also, the presence of $Z_2$ odd vector like singlet charged leptons can lead to dilepton plus MET signatures at colliders. As noted from the collider studies in \cite{Borah:2017dqx, Barman:2018jhz} within a similar model, the chosen values of vector like charged lepton masses are safe from existing collider limits. As can be seen from the above analysis, such bounds are satisfied by the parameter space shown in figure \ref{fig:model2a}, \ref{fig:model1a}, and \ref{fig:model1b}.
Apart from collider signatures of inert scalars in the model, as discussed in several earlier works mentioned above, the model can also have interesting signatures at lepton flavor violating decays like $\mu \rightarrow e \gamma, \mu \rightarrow e e e$ and $\mu \rightarrow e$ conversions \cite{Toma:2013zsa, Vicente:2014wga}. With inclusion of vector-like quarks and vector-like charged leptons, like we have in model-II, one can also explain flavor anomalies in $b \rightarrow s$ decays as well as muon $g-2$ \cite{Barman:2018jhz}. Thus, the model we have studied here, which successfully explains the ANITA anomalous events also have rich phenomenology that can be accessed at different experiments operating at cosmic, energy as well as intensity frontiers.

\section{Summary and Conclusions} 
\label{sec:concl}
We have proposed a beyond standard model framework to explain the two anomalous upward going ultra high energy cosmic ray air shower events reported by the ANITA collaboration. The novel feature of the model is the way it relates the origin of AAEs to non-thermal origin of dark matter in the universe as well as the origin of light neutrino masses. Sticking to minimality we extend the standard model by few additional particles, all of which are odd under an in-built and unbroken $Z_2$ symmetry. While two types of such $Z_2$ odd fields namely, a scalar doublet and gauge singlet neutral fermions play the role of generating light neutrino masses at one loop level similar to the scotogenic scenarios, the other fields are responsible for generating the AAEs. The non-thermal nature of dark matter is related to its tiny coupling with the next to lightest $Z_2$ odd particle whose long-livedness makes its passage through earth possible followed by its decay into DM and a tau lepton required to explain the AAEs.

We first consider a dark matter scenario where its relic is generated purely from non-thermal production. In this case, one of the three $Z_2$ odd gauge singlet neutral fermions is the DM candidate. Being gauge singlet, its non-thermal nature is dictated by its tiny coupling with the SM leptons and $Z_2$ odd scalar doublet. While such tiny coupling results in vanishingly small lightest neutrino mass \cite{Borah:2017dfn}, the abundance of DM is generated from decay of the $Z_2$ odd scalar doublet, the NLSP in this case. While the correct DM relic is obtained by suitable choices of DM, NLSP masses and their coupling, it was found that the model can not explain the AAEs. We then consider a hybrid setup where DM relic is obtained from both thermal as well as non-thermal contribution. The neutral component of a second $Z_2$ odd scalar doublet is considered to be the DM candidate which annihilates into the SM particles at a rate larger than the one required for correct thermal relic abundance. This leads to an under-abundant thermal dark matter and the deficit can be filled in by a non-thermal contribution from a singlet fermion, considered to be a charged vector like singlet fermion $E$. The light neutrino masses arise as usual from gauge singlet neutral fermions and another $Z_2$ odd scalar doublet, thereby not getting affected by the smallness of DM coupling with $E$. The model is found to explain the AAEs successfully.

Although we have confined our discussions here to DM relic and anomalous ANITA events, the Model-II of our study, successful in explaining both of these, can have other consequences which can be tested at experiments operating at different frontiers. For example, the components of $Z_2$ odd scalar doublet can be produced at a significant rate at collider experiments like the LHC due to its electroweak gauge interactions.  Our model also has additional colored vector like $Z_2$ odd fermions which can also give rise to similar signatures at colliders. While we briefly mention such additional aspects of probing this model at variety of experiments, the detailed phenomenological study of such scenarios is beyond the scope of this present work and left for future works.

\section*{Acknowledgements} 
DB acknowledges the support from IIT Guwahati start-up grant (reference number: xPHYSUGI-ITG01152xxDB001), Early Career Research Award from DST-SERB, Government of India (reference number: ECR/2017/001873) and Associateship Programme of Inter University Centre for Astronomy and Astrophysics (IUCAA), Pune. The research of UKD is supported by the Ministry of Science, ICT \& Future Planning of Korea, the Pohang City Government, and the Gyeongsangbuk-do Provincial Government. The research of GT is supported by the Basic Science Research Program through the National Research Foundation of Korea (NRF) funded by the Ministry of Education, grant number 2019R1F1A1052231. DB and UKD thanks the organizers of WHEPP XVI where important discussions are carried out. We would like to thank Bhavesh Chauhan for useful discussion.

%


\bibliographystyle{elsarticle-num}
\bibliography{ref_ftdm.bib}

\end{document}